\documentclass[a4paper,11pt]{article}
\usepackage{pos}

\newcommand{\ndat}{N_{\mathrm{dat}}}

\newcommand{\cov}{\mathrm{Cov}}
\newcommand{\FKtab}{(\mathrm{FK})}
\newcommand{\FKtabT}{(\mathrm{FK})^T}
\newcommand{\GP}{\mathcal{GP}}
\newcommand{\lat}{{\mathrm{lat}}}

\title{Inverse Problems in PDF Determinations}
\author[a]{Alessandro Candido}
\author*[b]{Luigi Del Debbio}
\author[c]{Tommaso Giani}
\author[d]{Giacomo Petrillo}

\affiliation[a]{
  TIF Lab, Dipartimento di Fisica, Universit\`a degli Studi di Milano and INFN
  Sezione di Milano \\
  Via Celoria 16, 20133 Milano, Italy
}

\affiliation[b]{Higgs Centre for Theoretical Physics, School of Physics \&
  Astronomy, The University of Edinburgh\\
  Peter Guthrie Tait Road, Edinburgh EH9 3FD, United Kingdom}

\affiliation[c]{Department of Physics and Astronomy, VU Amsterdam, 1081 HV Amsterdam, The Netherlands\\
\& Nikhef Theory Group, Science Park 105, 1098 XG Amsterdam, The Netherlands}

\affiliation[d]{Dipartimento di Statistica, Informatica, Applicazioni “Giuseppe Parenti” (DISIA), Universit\`a di Firenze \\
Viale Morgagni 59, 50134 Firenze, Italy}

\emailAdd{luigi.del.debbio@ed.ac.uk}
\abstract{The determination of Parton Distribution Functions from a finite set of data is a typical example of an inverse problem. Inverse problems are notoriously difficult to solve, in particular when a robust determination of the uncertainty in the result is needed. We present a Bayesian framework to deal with this problem and discuss first results from a closure test.}

\FullConference{%
  The 39th International Symposium on Lattice Field Theory (Lattice2022),\\
  8-13 August, 2022 \\
  Bonn, Germany 
}

\begin{document}
\maketitle

\section{Introduction}
\label{sec:Intro}

The determination of Parton Distribution Functions (PDFs) from lattice
simulations requires the solution of an inverse problem, where a function of a
real variable $f(x)$ needs to be reconstructed from a finite set of data $\{y_I;
I=1, \ldots, \ndat\}$. Following the original ideas in Ref.~\cite{Ji:2013dva},
the lattice data are correlators of fields, which are computed in Monte Carlo
simulations. Factorization theorems allow us to express these correlators as
convolutions of the PDFs,
\begin{equation}
  \label{eq:FactThm}
  y_I = \int_0^1 dx\, C_{Ij}(x) f_j(x)\, , 
\end{equation}
where $C_{Ij}(x)$ are Wilson coefficients that can be obtained in perturbation
theory, and the index $j$ is summed over all partons described by the PDFs; see,
e.g., Ref.~\cite{NNPDF:2021njg} for a recent determination of PDFs from
experimental data and Ref.~\cite{DelDebbio:2020cbz} for a (simple) discussion of
the factorization formula for lattice data in a toy model scalar field theory. 

We argued in Ref.~\cite{DelDebbio:2021whr} that a Bayesian approach is best
suited to characterise the knowledge about the function $f$, taking into account
any {\em prior} assumptions and the existing data. In that paper, the integral
in Eq.~\ref{eq:FactThm} was evaluated as a discrete sum on a grid of points
$x_j$,
\begin{equation}
  \label{eq:FactThmDiscrete}
  y_I = \sum_{j=1}^N \FKtab_{Ij} f(x_j)\, , \quad j=1, \ldots, N
\end{equation}
and a multivariate Gaussian with covariance $C_X$ was chosen as a prior for the
discrete set of values $f(x_j)$. Given the set of data, Bayes' theorem
provides the posterior distribution for the variables $f(x_j)$.

In this work, we are going to consider the `continuum limit' of this approach,
where the function $f$ is treated as a {\em Gaussian Process} (GP). GPs are
characterized by a mean function $m(x)$ and a covariance function $k(x,x')$, 
\begin{equation}
  \label{eq:GPDef}
  f \sim \mathcal{GP}(m,k)\, ,
\end{equation}
so that $x$ can be thought of as an index, and $f(x)$ as a stochastic variable indexed
by $x$. For any pair of indices $x$ and $x'$, the stochastic variables $f(x)$
and $f(x')$ are Gaussian variables, with 
\begin{align}
  \label{eq:GPMean}
  E[f(x)] &= m(x)\, , \\
  E[f(x')]&= m(x')\, , \\
  \cov[f(x),f(x')] &= k(x,x')\, .
\end{align}
Note that the kernel $k$, by determining the correlation between the values of
$f$ at different values of $x$, encodes somehow the smoothness of the function. 

\section{Inference Using Gaussian Processes}
\label{sec:Inference}

Before applying GPs to the determination of PDFs, let us briefly summarise their
general usage for inference purposes. When trying to determine a function $f$,
we are going to assume that the prior distribution of the function is given by a
GP, as specified in Eq.~\ref{eq:GPDef}. Let us now consider two sets of 
indices,  
\begin{equation}
  \label{eq:IndSets}
  \mathbf{x} = \begin{pmatrix}
    x_1 \\
    \vdots \\
    x_N
  \end{pmatrix}\, ,
  \quad
  \mathbf{x^*} = \begin{pmatrix}
    x^*_1 \\
    \vdots \\
    x^*_M
  \end{pmatrix}\, ,
\end{equation}
we can costruct two vectors, which contain the values of $f$ evaluated at the
points $\mathbf{x}$ and $\mathbf{x^*}$ respectively, 
\begin{equation}
  \label{eq:FvaluesIndSets}
  \mathbf{f}   = f(\mathbf{x})   \in \mathbb{R}^N\, , \quad 
  \mathbf{f^*} = f(\mathbf{x^*}) \in \mathbb{R}^M\, ,
\end{equation}
where the expressions above are a short-hand for $f_i=f(x_i)$ and
$f^*_i=f(x^*_i)$ respectively. The $(N+M)$-dimensional vector $\displaystyle
\begin{pmatrix} \mathbf{f} \\
  \mathbf{f^*}
\end{pmatrix}$ is a stochastic variable that is distributed according to the
prior Gaussian distribution, with mean~\footnote{The notation here and in 
Eq.~\ref{eq:CovIndSets} is a
generalization of the one explained in Eq.~\ref{eq:FvaluesIndSets}}
\begin{equation}
  \label{eq:MeanIndSets}
  \begin{pmatrix}
    \mathbf{m} \\
    \mathbf{m^*}
  \end{pmatrix}
\end{equation}
and covariance
\begin{equation}
  \label{eq:CovIndSets}
  \Sigma = \begin{pmatrix}
    k(\mathbf{x},\mathbf{x}^T) & k(\mathbf{x},\mathbf{x^*}^T) \\
    k(\mathbf{x^*},\mathbf{x}^T) & k(\mathbf{x^*},\mathbf{x^*}^T) 
  \end{pmatrix}
    = \begin{pmatrix}
      \Sigma_{xx} & \Sigma_{xx^*} \\
      \Sigma_{x^*x} & \Sigma_{x^*x^*}
    \end{pmatrix}\, .
\end{equation} 
When considering a vector $\mathbf{f}$ containing a finite number of variables,
this is exactly the formalism used in Ref.~\cite{DelDebbio:2021whr}; the only
difference being that, in the case of a GP, the mean and the variance are
dictated by the functions $m$ and $k$ that characterise the GP, as shown in
Eq.~\ref{eq:CovIndSets}. 

In what follows, we are going to consider two different examples where Bayesian
inference can be used. 
\begin{enumerate}
  \item We will first consider the case where the values of $\mathbf{f}$ are
  exactly known, and we will use this information to construct a posterior
  distribution for $\mathbf{f^*}$. We will refer to this case as {\em point-wise
  data}.

  \item We will then analyse the case where the function $f$ itself is not
  known, but we are given some data like the datapoints $y_I$ discussed above,
  i.e., data that is obtained as a linear transformation of the values
  $\mathbf{f}$. In this scenario, we will compute the posterior distribution for
  the values of $\mathbf{f}$ and $\mathbf{f^*}$ given the data $y_I$ and their
  covariance matrix. We will refer to this scenario as {\em lattice data}.
  
\end{enumerate}

\paragraph{Point-wise data.} 
It is interesting to remark that, because of the correlation between the values
of $f$ at different values of $x$, knowing the values of 
$\mathbf{f}$ yields some information on the values of $\mathbf{f^*}$. 
In a Bayesian framework, this information is extracted by
computing the posterior distribution for the stochastic variables
$\mathbf{f^*}$. Let us assume that the values of $f$ on the $\mathbf{x}$
index set are given by a vector $\mathbf{y}$,
\begin{equation}
  \label{eq:GPdata}
  \mathbf{y} = f(\mathbf{x})\, ,
\end{equation}
a standard calculation, using the Schur complement of $\Sigma_{xx}$, yields the
posterior distribution for the values of $f$ on the $\mathbf{x^*}$ set, 
\begin{equation}
  \label{eq:PostGP}
  \left(\mathbf{f^*} | \mathbf{f}=\mathbf{y}\right)
    \sim \mathcal{N}
    \left(\mathbf{\tilde m^*}, \tilde{\Sigma}_{x^*x^*}\right)\, ,
\end{equation}
where 
\begin{align}
  \label{eq:PostMean}
  \mathbf{\tilde m^*} &= \mathbf{m^*} + 
    \Sigma_{x^*x} \Sigma_{xx}^+ \left(\mathbf{y}-\mathbf{m}\right) \, , \\
  \tilde{\Sigma}_{x^*x^*} &= \Sigma_{x^*x^*} - 
    \Sigma_{x^*x} \Sigma_{xx}^+ \Sigma_{xx^*}\, ,
\end{align}
where $\Sigma_{xx}^+$ denotes the Moore-Penrose pseudoinverse of $\Sigma_{xx}$.

\paragraph{Lattice data.}
As discussed above, the lattice data, with their Monte Carlo covariance, are
typically expressed as linear functions of the PDFs. The knowledge of a variable
that depends linearly on the GP can be incorporated in our formalism by
introducing the stochastic variable
\begin{equation}
  \label{eq:EpsDistr}
  \mathbf{\epsilon} \sim \mathcal{N}\left(0, C_Y\right)\, ,
\end{equation}
and imposing that
\begin{equation}
  \label{eq:DataWithErr}
  \mathbf{y} = \FKtab \mathbf{f} + \mathbf{\epsilon}\, ,
\end{equation}
where $\mathbf{y}$ are the central values and $C_Y$ is the covariance matrix of
the lattice estimates. The linear dependence of $\mathbf{y}$ on $\mathbf{f}$ is
encoded in the matrix $\FKtab$. Note that in Eq.~\ref{eq:DataWithErr} we assume
that the observables, defined in Eq.~\ref{eq:FactThmDiscrete}, are computed
using the function $f$ evaluated on the points in $\mathbf{x}$. Prior knowledge
of the function and the Monte Carlo covariance of the data must be uncorrelated,
and therefore the covariance matrix of the three sets of stochastic variables
$(\mathbf{f},\mathbf{f^*},\mathbf{\epsilon})$ is a block-diagonal
$(N+M+\ndat)\times (N+M+\ndat)$ matrix
\begin{equation}
  \label{eq:ThreeVarsCov}
  \cov = \begin{pmatrix}
    \Sigma & 0 \\
    0 & C_Y
  \end{pmatrix}\, ,
\end{equation}
where $\Sigma$ is the $(N+M)\times (N+M)$ matrix introduced in
Eq.~\ref{eq:CovIndSets}.

Conditioning on the observed value $y$ in Eq.~\ref{eq:DataWithErr}, and
marginalizing $\mathbf{f^*}$ and $\mathbf{\epsilon}$, yields a Gaussian
posterior for $\mathbf{f}$,
\begin{equation}
  \label{eq:PostDataGauss}
  \left(\mathbf{f} | \FKtab \mathbf{f} + \epsilon = \mathbf{y}\right)
    \sim \mathcal{N}\left(\mathbf{\tilde m}, \tilde{\Sigma}_{xx}\right)\, , 
\end{equation}
with mean $\tilde{m}$ and covariance $\tilde{\Sigma}_{xx}$, given by 
\begin{align}
  \label{eq:MeanPostData}
  \mathbf{\tilde m} 
    &= \mathbf{m} + \Sigma_{xx} \FKtabT 
    \left(\FKtab \Sigma_{xx} \FKtabT + C_Y\right)^+ 
    \left(\mathbf{y} - \FKtab \mathbf{m}\right)\, ,\\
  \label{eq:CovPostDataBis}
  \tilde{\Sigma}_{xx} 
    &= \Sigma_{xx} - 
    \Sigma_{xx} \FKtabT \left(\FKtab \Sigma_{xx} \FKtabT + C_Y\right)^+ 
    \FKtab \Sigma_{xx}\, .
\end{align}
Note that Eq.~\ref{eq:CovPostDataBis} can equivalently be written as
\begin{equation}
  \label{eq:CovPostData}
  \tilde{\Sigma}_{xx}^{-1} 
    = \Sigma_{xx}^{-1} + \FKtabT C_Y^{-1} \FKtab\, .
\end{equation}
Eqs.~\ref{eq:MeanPostData} and~\ref{eq:CovPostData} were already obtained in
Ref.~\cite{DelDebbio:2021whr}, while Eq.~\ref{eq:CovPostDataBis} provides an
alternative expression for the posterior covariance. 

A similar derivation yields a multivariate Gaussian for the posterior
distribution of $\mathbf{f^*}$, 
\begin{equation}
  \label{eq:PostDataGP}
  \left(\mathbf{f^*} | \FKtab \mathbf{f} + \epsilon=\mathbf{y}\right)
    \sim \mathcal{N}
      \left(\mathbf{\tilde m^*}, \tilde{\Sigma}_{x^*x^*}^{\lat}\right)\, ,
\end{equation}
where in this case
\begin{align}
  \label{eq:PostMeanStar}
  \mathbf{\tilde m^*} &= \mathbf{m^*} + \Sigma_{x^*x} \FKtabT 
    \left(\FKtab \Sigma_{xx} \FKtabT + C_Y\right)^+ 
    \left(\mathbf{y} - \FKtab \mathbf{m}\right)\, ,\\
  \label{eq:PostCovStar}
  \tilde{\Sigma}_{x^*x^*}^{\lat} &= \Sigma_{x^*x^*} - 
    \Sigma_{x^*x} \FKtabT\, \left(\FKtab \Sigma_{xx} \FKtabT + C_Y\right)^+ \,
    \FKtab \Sigma_{xx^*}\, .
\end{align}
Introducing the corrections to the mean of the process due to Bayesian
inference,
\begin{align*}
  \label{eq:Delta1}
  \Delta\mathbf{m}   &= \mathbf{\tilde m} - \mathbf{m}\, , \\
  \Delta\mathbf{m^*} &= \mathbf{\tilde m^*} - \mathbf{m^*}\, , \\
\end{align*}
we have 
\begin{equation}
  \label{eq:MeanRelation}
  \Delta\mathbf{m^*} = \Sigma_{x^*x} \Sigma^+_{xx} \Delta \mathbf{m}\, ,
\end{equation}
and 
\begin{equation}
  \label{eq:CovRelation}
  \tilde\Sigma_{x^*x^*}^{\lat} = \tilde\Sigma_{x^*x^*} + \Sigma_{x^*x} \Sigma_{xx}^+
  \tilde\Sigma_{xx} \Sigma_{xx}^+ \Sigma_{xx^*}\, .
\end{equation}

\section{Hyperparameters}
\label{sec:HyperP}

In what follows, the mean value and the kernel of the GPs depend on a set of
hyperparameters that we denote by $\theta$, with their prior distribution,
$p_\theta(\theta)$. The $\theta$-dependent mean and kernel being henceforth
denoted $m_\theta(x)$ and $k_\theta(x,x')$, we define, in analogy with
Eq.~(\ref{eq:GPDef}),
\begin{equation}
  \label{eq:GPDefWithHyperP}
  \left(f|\theta\right) \sim 
    \GP\left(m_\theta, k_\theta\right)\, , \quad
    \theta \sim p_\theta\, .
\end{equation}
Every instance of the variable $\theta$ fully defines the process $f$. The
hyperparameters and their probability distribution become part of our Bayesian
analysis. The posterior distribution for the hy\-per\-parameters, given the
point-wise values $\mathbf{f}$, is simply
\begin{equation}
  \label{eq:HyperPPost}
  p(\theta | \mathbf{f}=\mathbf{y}) \propto 
    p(\mathbf{f}=\mathbf{y} | \theta) p_\theta(\theta)\, .
\end{equation}
Note that the first factor on the right-hand side of Eq.~(\ref{eq:HyperPPost})
is 
\begin{align}
  \label{eq:ProbFGivenTheta}
  p(\mathbf{f}=\mathbf{y} | \theta) = 
    \bigl[\det 2\pi \Sigma_{xx}(\theta)\bigr]^{-1/2} 
    \exp\left(
      -\frac12 (\mathbf{y}-\mathbf{m}(\theta))^T
      \Sigma_{xx}(\theta)^{-1}
      (\mathbf{y}-\mathbf{m}(\theta))
    \right)\, ,
\end{align}
where we have written explicitly the dependence of $\mathbf{m}$ and
$\Sigma_{xx}$ on the hyperparameters $\theta$. Similarly, mirroring the
discussion in the previous Section, the posterior probability of the
hyperparameters can be computed in the case where we know the central value and
the covariance of lattice data that are linear functions of the Gaussian
process,
\begin{align}
  \label{eq:HyperPPostLinearData}
  p\bigl(\theta | \FKtab \mathbf{f} + \epsilon=\mathbf{y}\bigr) &\propto 
    p\bigl(\FKtab \mathbf{f} + \epsilon=\mathbf{y} | \theta\bigr) p_\theta(\theta)\, , \nonumber \\
  &= \int d\mathbf{f}\, d\mathbf{\epsilon}\,
    p\bigl(\FKtab \mathbf{f} + \epsilon = \mathbf{y} | \mathbf{f}, \epsilon, \theta\bigr) 
    p(\mathbf{f}, \epsilon | \theta) p_\theta(\theta)\, .
\end{align}

We will also use the joint probability distribution of the values of the
function at the unseen points, $\mathbf{f^*}$ and the hyperparameters. In the
case of the point-wise data we obtain
\begin{align}
  \label{eq:JointPostValuesHyperPPoint}
  p(\mathbf{f^*}, \theta|\mathbf{f}=\mathbf{y}) 
    &= p(\mathbf{f^*}|\mathbf{f}=\mathbf{y},\theta) 
      p(\theta|\mathbf{f}=\mathbf{y})\, .
\end{align}
The first factor on the right-hand side has been computed in
Eq.~(\ref{eq:PostGP}), while for the second factor we can use the expression
above in Eq.~(\ref{eq:HyperPPost}). Finally, let us consider the case where the
data is the usual lattice data, as described in Section~2. The joint
posterior distribution in this case is
\begin{align}
  \label{eq:JointPostValuesHyperPLat}
  p\bigl(\mathbf{f^*},\theta | \FKtab \mathbf{f} + \epsilon=\mathbf{y}\bigr) 
    &= p\bigl(\mathbf{f^*}|\FKtab \mathbf{f} + \epsilon = \mathbf{y},\theta\bigr) 
      p\bigl(\theta|\FKtab \mathbf{f} + \epsilon = \mathbf{y}\bigr)\, .
\end{align}
For completeness, we remind the reader that the first factor on the right-hand
side has been computed in Eq.~\ref{eq:PostDataGP}, while the second factor is in Eq.~\ref{eq:HyperPPostLinearData}.

\section{Numerical Implementation}
  \label{sec:NumImpl}

In these proceedings, we are going to focus on some preliminary `closure tests'
of the method, where we generate artificial data from some known, input PDFs and
then use Bayesian inference to reconstruct the PDFs using the data. At this
stage, the input PDFs do not need to be realistic as we are mainly interested in
testing the methodology. We are going to use a set of these artificial PDFs in
the evolution basis~\footnote{see e.g. Ref.~\cite{NNPDF:2021njg} for their
definition in terms of the PDFs in the flavor basis. The singlet PDF is
traditionally denoted by $\Sigma$ and should not be confused with the covariance
matrix introduced above.}, 
\begin{equation}
  \label{eq:EvolBasis}
  f_k=\left\{V, V_3, V_8, T_3, T_8, T_{15}, \Sigma, g\right\}\, ,
\end{equation}
to generate 3000 data points using a fixed set of FK tables. These data are then
used to constrain the GPs, as dictated by the inference framework described
above. The result of the inference procedure can then be compared to the input
PDFs in order to assess the effectiveness of the methodology. 

Following the prescription in Section~\ref{sec:Inference}, we associate to each
PDF a GP with zero mean and Gibbs kernel~\cite{Rasmussen2006}
\begin{equation}
  \label{eq:GibbsKern}
  k(x,x') = \sigma^2 \sqrt{
    \frac{2 \ell(x) \ell(x')}{\ell(x)^2 + \ell(x')^2} 
  } \exp \left[
    -\frac{\left(x-x'\right)^2}{\ell(x)^2 + \ell(x')^2}
  \right]\, ,
\end{equation}
where 
\begin{equation}
  \label{eq:EllDef}
  \ell(x) = \ell_0 (x+\varepsilon)\, ,
\end{equation}
$\sigma$ and $\ell_0$ are hyperparameters and $\varepsilon>0$ regularizes the
singularity at $x=0$. 

\paragraph{Physical Constraints.} Since the functions $T_i$ do not satisfy any
particular constraint, we associate a GP to each of them
\begin{equation}
  \label{eq:TGP}
  T_i \sim \GP\left(0,k\right)\, .
\end{equation}
On the other hand, the valence PDFs need to satisfy the constraints dictated by
the valence sum rules: 
\begin{align*}
  \int_0^1 dx\, V(x) &= 3 \, , \qquad
  \int_0^1 dx\, V_3(x) = 1 \, , \\
  \int_0^1 dx\, V_8(x) &= 3 \, , \qquad
  \int_0^1 dx\, V_{15}(x) = 3 \, .
\end{align*}
In order to implement these constraints, we associate a GP to the indefinite
integrals of $V_i$, denoted as $\tilde{V}_i$, so that
\begin{equation}
  \label{eq:VGP}
  V_i(x) = \tilde{V}_i'(x)\, , \qquad 
  \tilde{V}_i \sim \GP(0, \kappa)\, ,
\end{equation}
where $\kappa(x,x')=\ell(x) k(x,x') \ell(x')$. 
The sum rules can be expressed as linear constraints, 
\begin{align*}
  \tilde{V}(1) - \tilde{V}(0) &= 3\, , \qquad
  \tilde{V}_3(1) - \tilde{V}_3(0) = 1\, , \\
  \tilde{V}_8(1) - \tilde{V}_8(0) &= 3\, , \qquad
  \tilde{V}_{15}(1) - \tilde{V}_{15}(0) = 3\, . 
\end{align*}
The momentum sum rule, 
\begin{equation}
  \label{eq:MomSumRule}
  \int_0^1 dx\, x \left(\Sigma(x) + g(x)\right) = 1\, ,
\end{equation}
is implemented by defining
\begin{align*}
  x \Sigma(x) &= \tilde{\Sigma}'(x)\, , &
  x g(x) &= \tilde{g}'(x)\, , \\
  \tilde{\Sigma}(x) &= \frac{x^{a_\Sigma+1}}{a_\Sigma+2} 
    \tilde{\tilde{\Sigma}}(x)\, , &
  \tilde{g}(x) &= \frac{x^{a_g+1}}{a_g+2} 
    \tilde{\tilde{g}}(x)\, , \\
  \tilde{\tilde{\Sigma}} &\sim \GP(0,k)\, , &
  \tilde{\tilde{g}} &\sim \GP(0,k)\, ,
\end{align*}
and requiring
\begin{equation}
  \label{eq:MomSumRule}
  \tilde{\Sigma}(1) + \tilde{g}(1) -
    \tilde{\Sigma}(0) - \tilde{g}(0) = 1\, .
\end{equation}
In choosing this particular parametrization, we have introduced two
hyperparameters $a_\Sigma$ and $a_g$ that control the asymptotic behaviour of
$x\Sigma(x)$ and $xg(x)$ at small $x$. Since the PDFs vanish at $x=1$, we have
one final set of constraints
\begin{equation}
  \label{eq:VanishAtZero}
  \Sigma(1) = g(1) = V_i(1) = T_i(1) = 0\, .
\end{equation}

\paragraph{Workflow.} The full workflow for the closure test is as follows. 

\begin{itemize}
  \item The input PDFs are generated by sampling the prior distribution for the hyperparameters, and then sampling the GPs. 

  \item The artificial data are obtained from Eq.~\ref{eq:DataWithErr}, using a
  fixed set of FK tables. Each data point is assigned an independent error equal
  to 10\% of the range of the data points. In this particular examples the FK
  tables are such that points with $x<10^{-4}$ do not contribute to the
  observables. 

  \item The hyperparameters are fitted by computing the maximum of their
  posterior distribution. 
  
  \item Using the fitted value for the hyperparameters, we compute the posterior
  mean and covariance of the GP for each PDF. These posterior distributions -- characterized by their central values and covariances -- are what we call `{\it fitted PDFs}' in this framework. 
  
  \item The result of the Bayesian inference for the data points is computed
  using the fitted PDFs and the FK tables. 
\end{itemize}

\begin{figure}[ht!]
  \begin{center}
    \includegraphics[scale=0.5]{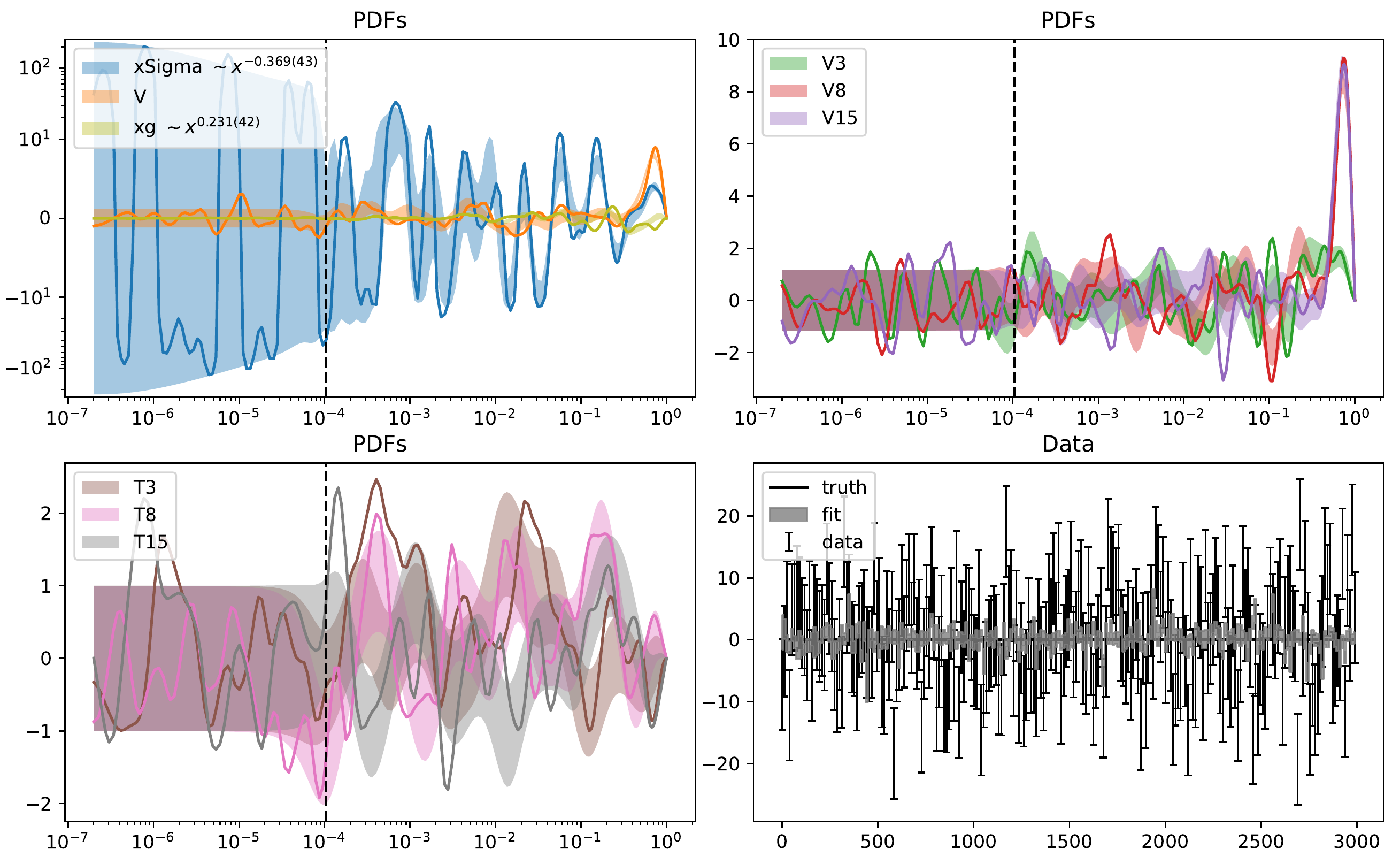}    
  \end{center}
  \caption{Results of the closure test. The lines represent the PDFs used to generate the artificial data, while the bands correspond to the $\pm 1\sigma$ intervals of the posterior distribution. The value of the PDFs at points to the left of the vertical dashed lines do not enter in the observables. The bottom right plot shows the difference between the artificial data set used for the inference and the theoretical prediction that is obtained from the posterior distribution of the GPs.   \label{fig:ClosureRes}
  }
\end{figure}

\paragraph{Results.} The results of this closure test are summarised in
Fig.~\ref{fig:ClosureRes}. The plots show a generic agreement between the input
PDFs, represented by the coloured lines, and the results of the Bayesian
inference, which are reported in the figure by drawing the $\pm 1\sigma$
interval at each point. As expected, the points for $x<10^{-4}$ are less
constrained by the data since they do not influence the data points. The data
points themselves are well reproduced by the posterior of the GP, as shown in
the bottom right plot. Some discrepancies in the PDFs are visible, e.g., for
$T_3$. A more quantitative analysis is postponed to future works. 

\section{Conclusions}

In these proceedings, we have sketched the application of GPs for the solution
of inverse problems like the ones that appear in the determination of PDFs from
data. Our methodology can be applied to both the cases of lattice data and
experimental data. It provides a mathematically robust tool to reconstruct the
PDFs while taking into account statistical errors and prior knowledge in a
robust mathematical formulation. A detailed study of the applicability of this
method, and its potential sources of bias, requires a detailed, systematic
investigation. 

The fact that an inifinitely-wide neural network behaves like a GP suggests that
there is rigorous connection between the methodology presented here and in
Ref.~\cite{DelDebbio:2021whr} and the fits based on neural network
parametrizations that have been in use for many decades. These features are
under investigations and we hope to report more results soon.

\end{document}